\RequirePackage{fix-cm}
\documentclass[smallextended]{svjour3}       
\smartqed  
\usepackage{graphicx}
\begin{document}

\title{A study of crossover from 3D ferrimagnetic Bulk $NiCr_{2}O_{4}$ compound into 2D spin-glass like nanophase.
}

\titlerunning{2D spin-glass phase in $NiCr_{2}O_{4}$ nanoparticles }        

\author{H. Singh, T. Ono, T. Chakraborty, K Srikanth, A. Venimadhav, R. Chandra, C. Mitra \and U. Kumar* 
}

\authorrunning{H. Singh} 

\institute{H. Singh, T. Chakraborty, K. Srikanth, C. Mitra, U. Kumar \at
              Indian Institute of Science Education and Research (IISER) Kolkata, Mohanpur Campus, PO: BCKV Campus Main Office, Mohanpur – 741252, Nadia, West Bengal, India. \\
              Tel.: +91-33-25873121\\
              Fax: +91-33-25873020\\
              \email{udayphy@iiserkol.ac.in}           
           \and
           T. Ono \at
              Department of Physical Science, Osaka Prefecture University, Gakuen-cho 1-1, Naka-ku,
Sakai, Osaka 599-8531, Japan.
			\and
			A. Venimadhav \at
			Cryogenic Engineering Centre, Indian Institute of Technology, Kharagpur-721302, India.
			\and
			R. Chandra \at
			Nano Science Laboratory, Institute Instrumentation Centre and Centre of Nanotechnology, 
Indian Institute of Technology Roorkee, Roorkee 247667, Uttarakhand, India.
}

\date{Received: date / Accepted: date}

\maketitle

\begin{abstract}
In this report, the magnetic behaviour of $NiCr_{2}O_{4}$ bulk and nanoparticle samples under different applied magnetic field has been investigated extensively. Nanoparticles of $NiCr_{2}O_{4}$ were obtained by mechanical milling of polycrystalline powder prepared by polyol method. FC-ZFC measurement of bulk at different applied magnetic field has revealed the existence of a ferrimagnetic transition around 66K followed by an antiferromagnetic transition close to 30K. However, its nano counterpart has shown remarkable change in magnetic properties - a suppression of ferrimagnetic transition accompanied by  strengthening low temperature magnetic phase and observation of a new transition at 90K ($T_P$), which is weakly magnetic in nature. The frequency dependent ac susceptibility data of nanoparticle have been fitted to the well known de Almedia-Thouless equation and a $H^{2/3}$ dependence of the low temperature peak is observed with a resulting zero field freezing temperature ($T_f^0$) equal to 10.1K. Further, the dynamical behaviour near freezing temperature has been analysed in terms of critical behaviour and the  obtained fitted parameters values being as $\tau_0$(relaxation time constant) = $3.6 X 10^{-6}s$, $T_f^0=8.7$K and $z\nu = 11.1$. Moreover, Vogel-Fulcher law has been used to understand the nature of freezing transition and the parameter after fitting are obtained as $E_a/k_B = 58.9$K, $\tau_0 = 5.22 \times 10^{-8}$ and $T_0 = 8.03$K. Finally, the spin-glass phase is concluded. Moreover, in contrast to bulk, the $H^{2/3}$ dependence of freezing temperature of nanoparticle sample (75h) does support the 2D surface like spin glass nature.
\end{abstract}

\section{Introduction}
One of the interesting family of magnetic materials is spinel oxides, mainly represented by the formula $AB_{2}O_{4}$ \cite{krupika}. Here the tetrahedral `A' sites were occupied by divalent ion and octahedral `B' site were equally occupied by divalent and trivalent ions. Among normal spinel compounds, nickel chromite has recently received considerable interest because of its classification into dynamically spin frustrated  systems, based upon the realization of resonance like magnetic excitation observed in neutron scattering experiment \cite{tomiyasu_77}. However, despite the presence of pyrochlore lattice which is formed by $Cr^{3+}$ ion, the occurrence of Jahn-Teller distortion nullify the possibility of $NiCr_{2}O_{4}$ to be considered as geometrically frustrated magnet. A number of potential application (light or heat sensitive micromechanical device, catalytic materials, gas sensors) has provided technological recognition to this material \cite{klemme,crottaz}.
Single crystals of $NiCr_{2}O_{4}$ has been extensively studied from the perspective of neutron scattering \cite{tomiyasu_77}, heat capacity \cite{klemme}, magnetodielectric \cite{mufti} and magnetic \cite{tomiyasu} measurements. However, magnetic studies of nanoparticles of this compound have not been reported so far. 
Below a critical physical dimension (in nanometer range) magnetic nanoparticles become single domain as compared to the normal multi-domain structure of the bulk counterpart. These single domain nanoparticle are quite interesting owing to unusual phenomenon they exhibit like superparamagnetism \cite{tejada1}, quantum tunnelling of magnetization \cite{tejada2} and large coercivities \cite{kneller}. The magnetic studies investigated have shown the occurrence of two transition temperatures $T_C$ and $T_S$, corresponding to the onset of ferrimagnetic (longitudinal) component and spiral (transverse) antiferromagnetic component. Recently , Tomiyasu \textit{et. al.} has reported a new magnetic structure in which for transverse and longitudinal components, B-sites were sorted into two sublattices \cite{tomiyasu}. $NiCr_{2}O_{4}$ undergoes structural transition from cubic to tetragonal structure below 310K owing to Jahn-Teller distortion. The $\frac{c}{(c-a)}$ ratio is 4\% at 4.2K \cite{prince}. However, Ishibashi \textit{et. al.} has confirmed another structural transition to orthorhombic structure at 65K and is correlated with the onset of ferrimagnetic ordering \cite{ishibashi}.
In the present study, we reported the magnetic study of $NiCr_{2}O_{4}$ nanoparticles prepared by mechanical milling of polycrystalline powder. FC-ZFC measurement were done at 0.005 and 0.1T applied magnetic field to investigate the effect of magnetic field on the behaviour of  the nanoparticle samples. The observed magnetic transition are in good agreement with the earlier reported results. However, a new transition ($T_P$), at 90K has been reported for the first time in bulk $NiCr_{2}O_{4}$. In high field measurement this transition disappears. But in nanoparticle samples, softening of $T_C$ with decrease in particle size was found, whereas $T_P$ is distinctly visible. The decrease in magnetization values in the magnetic nanoparticle system in comparison with its bulk phase is well known but still the subject matter of debate. This could be seen in the present study. The reason for this reduction in magnetization could be associated with canted spin arrangement or  spin disorder on the surface or finite size effect \cite{coey,morrish,parker}. The result of these phenomenon leads to superparamagnetism or spin-glass phases. We have explored the low temperature regime (down to 0.4K) as a function of magnetic field to investigate in detail the magnetic behaviour of the low temperature anomaly ($T_S$). Frequency dependent ac susceptibility was also carried out to reveal any frequency dependent magnetic phase. A 2D spin-glass like behaviour has been established from de Almeida-Thouless(AT) analysis which is further supported by frequency dependent ac susceptibility study. Finally, we tried to understand the mechanism responsible for the existence of 2D spin glass like phase and the surface spin disorder is concluded as the reason behind such an observation.

\section{Experimental Details}
$NiCr_{2}O_{4}$ bulk was prepared by decomposition of $NiCr_{2}O_{4}$ obtained by reacting $NiCl_{2}.6H_{2}O$ with $(NH_{4})2.Cr_{2}O_{7}$. In a typical preparation, 10 mmol of $NiCl_{2}.6H_{2}O $ and 10 mmol of $(NH_{4})2Cr_{2}O_{7}$ were mixed in 40 mL distilled water. To this mixture, 90 mmol of ethylene glycol mixed in 40 mL of distilled water was added. The resulting solution was stirred for 12 hours after which the solution was heated at 60°C to evaporate water and obtain a brown coloured gel which on further heating results in a greenish-black powder precursor. The greenish-black powder was heated at 350°C to remove ethylene glycol and ammonium chloride which was finally heated at 1200°C for 72 hours to obtain $NiCr_{2}O_{4}$ bulk crystalline form. This bulk powder form was used for the synthesis of different nano size $NiCr_{2}O_{4}$ sample using high energy ball mill machine Fritsch Planetary Mono Mill Pulverisette 6. For this purpose, the bulk powder sample was kept in an 80 ml agate bowl with 10 mm agate ball in 1: 8 sample and ball weight ratio and milled up to different milling time. All the nano size samples were annealed at 400°C for 4 hours and cooled very slowly to room temperature (40°C/h) to reduce the possible existing strain in the nano size sample. The phase purity and crystal structure (from 320 K to 10 K) of all the powder samples were studied by using Rigaku’s SmartLab X-ray diffractometer using Cu Kα lines in parallel beam geometry mode.

The X-ray diffraction (XRD) pattern of all the samples were scanned from 15 to 70° at the step angle of 0.02° and anode power 9 kW. To know the exact particle size distribution, the samples were subjected to transmission electron microscopy study and micrographs were collected under 200 kV anode voltage using model. A set of three sample 0 (Bulk), 33, 75h were prepared and subsequently studied. Zero field cool (ZFC) and field cool (FC) temperature dependent magnetic moment measurement was done at 0.005 and 0.1T field using Magnetic Property Measurement System (MPMS).  A systematic minimization of the trapped magnetic field in the superconducting coil of MPMS, is usually performed before commencing any measurement. To study the low temperature transition, ZFC measurement up to very low temperature of 0.4K were carried out at applied magnetic field of 0.005, 0.01, 0.05 and 0.1T. These low temperature measurements were performed at Osaka, Japan. The frequency dependent ac susceptibility curves as a function of temperature were also collected at 7, 73 and 143 Hz in the MPMS at small oscillating magnetic field of 1Oe amplitude.

\section{Results and Discussion}
\begin{figure}[t]
\begin{center}
   \includegraphics[width=6.0cm]{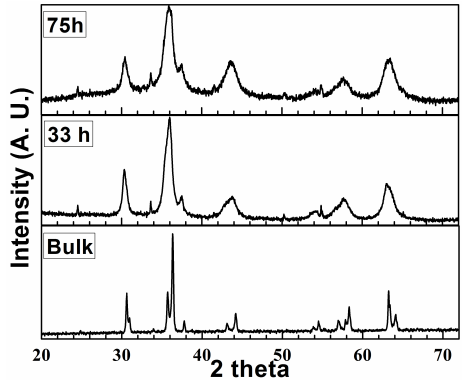}
\end{center}
 \caption{X-ray pattern of bulk and nanoparticle samples prepared with milling time of 33 and 75h, shown along with bulk sample.}
 \label{Fig1}
 \end{figure}

\begin{figure}[t]
\begin{center}
   \includegraphics[width=4.5cm]{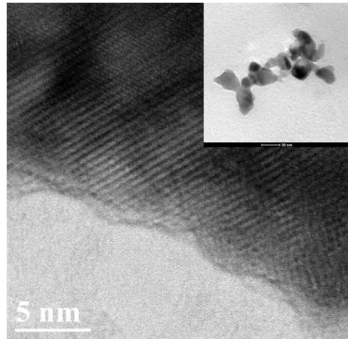}
\end{center}
 \caption{TEM image of the 75h milled sample revealing the lattice grating lines. In the inset is the particle size distribution of the same sample with average particle size close to 12nm .}
 \label{Fig2}
 \end{figure}

The room temperature XRD pattern of bulk, 33h, and 75h milled samples are shown in Fig. \ref{Fig1}. The diffraction pattern of the bulk $NiCr_{2}O_{4}$ powder sample is identical to the earlier reported results \cite{mufti} confirming the phase purity of the sample. The profile fitting for bulk diffraction patterns was carried out using Fullprof Suite version 2009 and structural parameters were determined. Above 310 K, the bulk $NiCr_{2}O_{4}$ exits in cubic spinel structure with space group Fd(-3m) and lattice parameter 8.3194\AA. However, it adopts a tetragonal cubic spinel structure with space group I 41/amd and lattice parameters of a = b = 5.8369 \AA  and c = 8.4312\AA  below 310 K. The splitting of Bragg peaks in Fig. \ref{Fig1} can be seen as an indication of cubic tetragonal structure and is because of Jahn-Teller distortion. These lattice parameters for cubic and tetragonal structure are in good agreement with earlier reported values \cite{mufti,tomiyasu}. A clear and systematic base broadening in the XRD pattern of 75h and 33h samples compared to bulk one can be seen in Fig. \ref{Fig1} which is associated to decreasing particle size effect. The dominant peak at nearly 36° [(211)d plane] is taken under consideration for particle size determination using standard Debye-Scherrer equation. The average particle size of 33h and 75h ball milled samples was 14.5 and 22.15 nm respectively. To confirm these particle sizes, the 75h sample were subjected to TEM study as shown in Fig. \ref{Fig2}. Distinctly visible lattice fringes confirms the high quality of the sample as is usually observed in single crystalline nanostructures.

\begin{figure}[t]
\begin{center}
   \includegraphics[width=8.5cm]{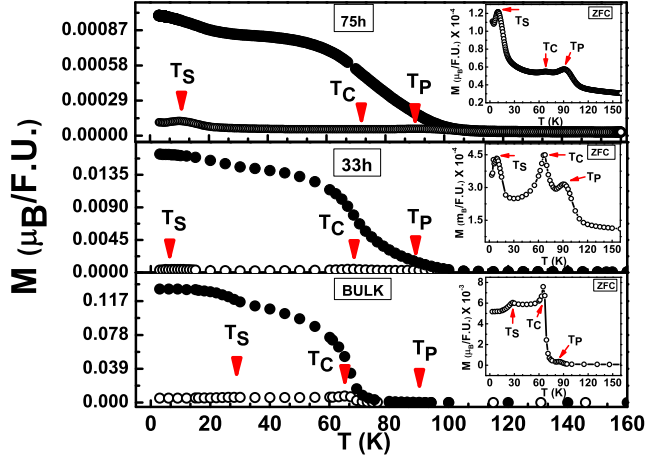}
\end{center}
 \caption{Magnetization FC-ZFC measurement of 75h, 33h and bulk  $NiCr_{2}O_{4}$ sample in 0.005T cooling field. Inset shows the ZFC of respective samples in 0.005T cooling field. All the three transitions can be clearly seen.}
 \label{Fig5}
 \end{figure}

FC-ZFC measurement of bulk and nanoparticle samples were performed at different applied magnetic field (0.005, 0.01, 0.1 and 2T). For 0.005T applied magnetic field, in case of bulk sample, the observed transitions at 67K and 29K corresponding to $T_{C}$ and $T_{S}$ is well reproducible  and is close to the reported results in this system \cite{tomiyasu} as shown in Fig. \ref{Fig5}. We have zeroed the magnetic field of the superconducting magnet of MPMS by following a standard zeroing protocol to ensure that the field was zero at the time of cooling. Interestingly, the ZFC measurement at 0.005T of this system has shown a new transition at 90K (named as $T_{P}$ here) as shown in the inset of Fig. \ref{Fig5}. ZFC measurement at 0.005T has not been reported so far in the literature for this system. We argue that there is development of a weak magnetic phase in the system at this transition temperature, however, the real nature of interaction might be more complex. Study performed by Ishibashi et. al., has shown that the high temperature magnetic transition is occurring simultaneously with the structural transition from tetragonal to orthorhombic at $T_{C}$ \cite{ishibashi}. The temperature dependent XRD measurement (not shown here) performed over bulk sample did not revealed any structural transition happening at 90K. Hence, we verified that this new transition at $T_{P}$ is solely magnetic.

\begin{figure}[t]
\begin{center}
   \includegraphics[width=8.5cm]{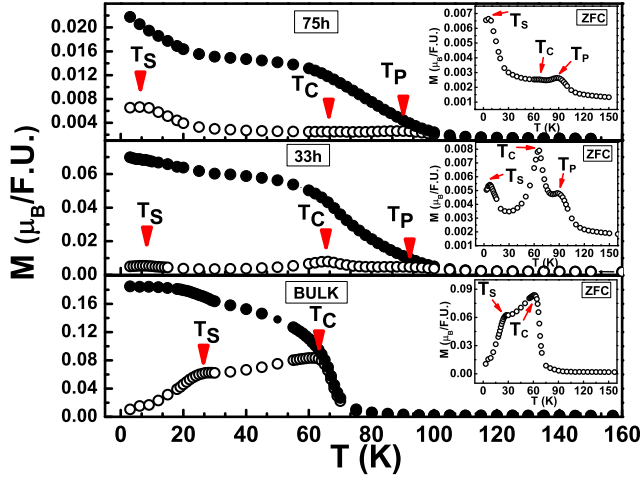}
\end{center}
 \caption{FC-ZFC magnetization measurement of 75h, 33h and bulk  $NiCr_{2}O_{4}$ sample in 0.1T cooling field. Inset shows the ZFC of respective samples. All the three transitions can be clearly seen.}
 \label{Fig1K}
 \end{figure}

The variation of temperature dependent magnetic behaviour at 0.005T applied magnetic field for all the samples is shown in Fig. \ref{Fig5}. The decrease in magnetization at  $T_{C}$ is observed, with decrease in particle size. This softening of $T_{C}$ indicates the breaking of spin-lattice coupling owing to particle size reduction. The presence of spin-lattice coupling is common among chromium based spinels. The weakening of spin-lattice coupling is also supported by a shift in of spiral AFM ordering at $T_{S}$ in our samples towards lower temperature, as the particle size decreases and can be seen in the insets of Fig. \ref{Fig5} \cite{ueda}. The decrease in particle size destroys the long range AFM correlation, resulting in a shift in $T_S$ towards lower temperatures \cite{chen}. As the particle size decreases, a reduction in the magnetization value is observed. Lowering of magnetization can be explained through increased surface spin canting, spin disorder or finite size effect at nanoparticle surface \cite{kodama}. All the samples were also investigated at 0.01T magnetic field as shown in Fig. \ref{Fig1K}. Apart from increase in magnetization value no significant effect of magnetic field is observed. However, the new transition $T_P$ could not be observed distinctly. This could be due to the formation of ferrimagnetic spin clusters which is overcome by the application of higher fields causing a disappearance of $T_P$ at higher fields. On the other hand, $T_P$ is clearly visible in the nanoparticle samples. The exchange coupling between magnetic ions ($Ni^{2+}$ and $Cr^{3+}$) is directly proportional to Curie-Weiss temperature ($\theta_{CW}$) and can be expressed by relation,

\begin{equation}
J = A \vert \theta_{CW} \vert
\label{cw}
\end{equation}

where, $A = \frac{3k_{B} }{ZS(S+1)}$, $k_{B}$ is the Boltzmann constant, Z is the number of nearest neighbour interaction and S is the total spin. The high temperature region ($T>150K$) of ZFC magnetization data at 0.005 and 0.01T was fitted with the Curie-Weiss law and subsequently $\theta_{CW}$ value was calculated for both bulk and nanoparticles. For 0.005T field value, the $\theta_{CW}$ for bulk, 33 Hrs and 75h samples are found to be $-346.6\pm 2, -295.05\pm2$ and $-238.26\pm2$K respectively. However, for 0.01T field value, it varies as $-607.23\pm2, -295.4\pm2$ and $ -236.3\pm2$K. The negative value of $\theta_{CW}$ indicates the presence of antiferromagnetic correlation interaction. One can see that for both 0.005 and 0.01T field, $\theta_{CW}$ decreases with decrease in particle size. Now, according to eq. \ref{cw}, with decrease in $\theta_{CW}$, the exchange interaction among magnetic ions decreases and this signifies the weakening of $T_C$ in our nanoparticle samples. According to mean field theory the frustration parameter is defined as f = $\theta_{CW} /T_N$. For bulk, `f' is nearly 20 and for 75h nanoparticle sample it is 23.6. The general condition for the existence of geometrically frustrated magnets, the ground state should be either antiferromagnetic or spin-glass in nature \cite{ramirez}.  

\begin{figure}[t]
\begin{center}
   \includegraphics[width=8.5cm]{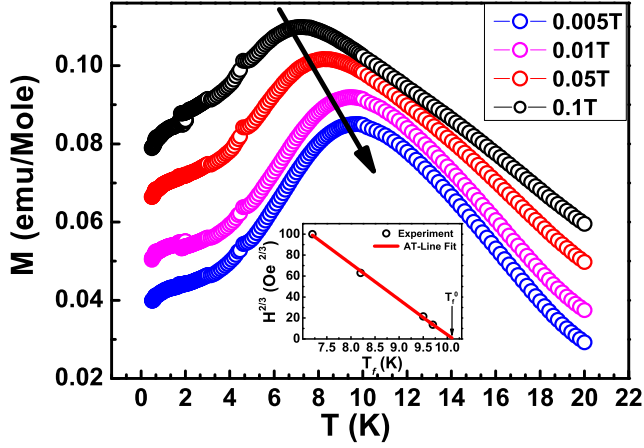}
\end{center}
 \caption{ZFC curves for 75h sample, measured at different cooling field of 0.005, 0.01, 0.05, 0.1T. The inset shows the field field dependence of freezing temperature and the solid line represents a fit with De- Almedia Thouless equation. }
 \label{FigHe_rrr}
 \end{figure}

To investigate the low temperature behaviour of $T_S$, magnetization measurement down to 0.4K at different applied dc magnetic fields were carried out as shown in Fig. \ref{FigHe_rrr}. With increase in applied dc magnetic field, the peak temperature ($T_S$) is found to be shifted to lower temperatures which is a characteristic feature of a spin-glass. In such a situation, the peak temperature ($T_S$) is identified as the freezing temperature represented as $T_f$ in the ensuing analysis.  The small kinks seen in the main panel of Fig. \ref{FigHe_rrr}, in low temperature region ($T < 6K$) are may be due to experimental errors. We fit the observed freezing temperature to the power law relation which is well known as De Almeida–Thouless (AT) equation \cite{at}, given by

\begin{equation}
\frac{H_{AT}(T_f)}{\Delta J} \propto \left( 1-\frac{T_f}{T^0_f}\right)^{3/2}
\label{at}
\end{equation}

and is shown in the inset of Fig. \ref{FigHe_rrr}. Here, $\Delta J$ is the width of the distribution of exchange interaction and $T_f^0$ is the freezing temperature at zero magnetic field. The fitting parameter we used is $T_f^0$. The zero field freezing point was obtained by extrapolating the AT line to the temperature axis and is found to be $T_f^0$=10.1K as shown in the inset of Fig. \ref{FigHe_rrr}. The error in determining freezing temperatures is very small (barely visible) and is evaluated by the temperature step size for the measurement of $\sim$ 0.2K.  It can be seen that the freezing temperature $T_f$, corresponding to which the magnetization value $M_{FC}-M_{ZFC}=\Delta M$ becomes non-zero (Fig. \ref{Fig5},   \ref{Fig1K}) indicating the onset of freezing temperature, decreases (low temperature shift) with increase in applied dc magnetic field (H). It shows a $T_f \propto H^{2/3}$ dependence, which reflects 2D surface like spin-glass behaviour. It is worth mentioning here that, in bulk the system reflects 3D ferrimagnetic behaviour and in nanoparticle regime it reflects a 2D spin-glass like behaviour with $T_f = 10.1K$. This 2D spin-glass like nature is associated with the surface spin character of the $NiCr_2O_4$ nanoparticles. However, core of the nanoparticle is still ferrimagnetic in nature. In this regard, the temperature dependent response of magnetization from bulk to nanoparticle could be more informative (Fig. \ref{Fig5}, \ref{Fig1K}). The weakening of ferrimagnetic transitions ($T_C$) and strengthening of low temperature peak ($T_N/T_f$) could be seen from bulk to nanoparticle regime with the evolution of a new peak ($T_P\sim90K$) towards higher temperature side. Obviously, the surface (shell) spins have dominating role/effect compared to bulk (core) spins in the nanoparticles which leads to the conclusion that ferrimagnetically ordered spins are quite less in number in comparasion to the surface spins. The origin and nature of the $T_P$ will be discussed latter in the letter. 

\begin{figure}[t]
\begin{center}
   \includegraphics[width=8.5cm]{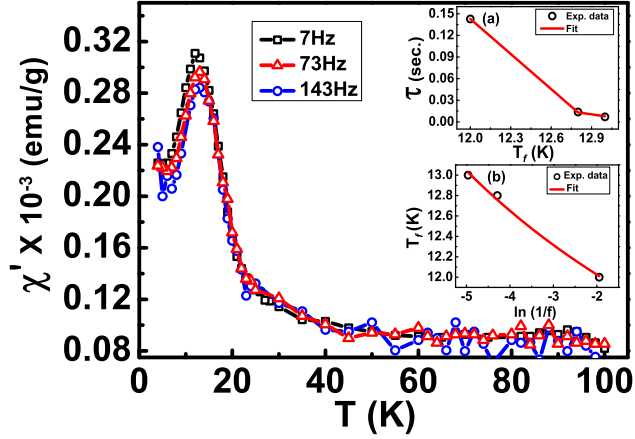}
\end{center}
 \caption{A plot of AC susceptibility as a function of temperature measured at different frequencies of 7, 73 and 143 Hz for 75h sample. The inset (a) and (b) shows the fitting as per eq. \ref{plaw} and \ref{vf law}.}
 \label{chi}
 \end{figure}

The presence of spin-glass like phase in the reported system might be due to superparamagnetism and AT line analysis alone is not sufficient. Hence it is necessary to implement another experimental evidence to establish the claim of spin-glass phase in our system. Thus, the frequency dependent ac susceptibility analysis was done to establish this claim.
Fig. \ref{chi}, shows the real part of the ac susceptibility ($\chi'$) as a function of temperature measured at frequencies 7, 73, and 143 Hz. For this measurement, the sample was first cooled down to 4K from 300K. Then a probing ac magnetic field of amplitude 1 Oe was applied for the ac susceptibility measurement. The dc biasing field was set to zero during data accusation. The freezing temperature ($T_f$) is identified as the peak in the curve and found to be shifted towards high temperature with increase in frequency. It is believed that superparamagnetism (SPM) can also give rise to peaks in ac `$\chi$' measurements accompanied by frequency dependent shift in peak (blocking temperature in SPM) positions can be seen, similar to the results shown in Fig. \ref{chi}. The two magnetic phases (SPM and spin-glass) can be distinguished by the empirical quantity $\Delta T_f/[T_f log(f)]$, with values varying from 0.004-0.018 for spin-glass to as large as 0.3 for SPM \cite{mydosh}. Here, $\Delta T_f = T_f-T_f^0$, is the shift in freezing temperature for spin-glass phase from $T_f^0$.  For 75h sample, the value of the empirical quantity is found to be 0.05 which is close to the spin-glass phase. To understand the dynamical behaviour of the spin glass phase near freezing temperature, two different theoretical approaches have been adopted. The first approach assumes that a phase transition takes place at the freezing temperature in the vicinity of which a critical behaviour in the temperature dependence of relaxation time ($\tau$) is expected. 
This is known as critical slowing down model. In this model, the critical relaxation time, defined as $\tau=1/f$, near the transition point is related to the correlation length ($\frac{T_f}{T^0_f}-1$) of the spins and is expressed in the form of a power law as given below \cite{souletie,edwards,mukadam}. The freezing temperatures obtained from Fig. \ref{chi} were fitted with the power law equation,
\begin{equation}
\tau=\tau_{0} \left(\frac{T_{f}}{T^{f}_{0}}-1\right)^{-z\nu},
\label{plaw}
\end{equation}
where, $\tau$=$1/f$, f is the frequency of the ac $\chi$ measurement, $\tau_0$ is the relaxation time constant normally lying in the range of $10^{-9}$ to $10^{-13}$ sec and $z\nu$ is the critical exponent. The fitting shown in the inset (a) of Fig. \ref{chi}, has yielded parameter values as $\tau_0 = 3.6 \times 10^{-6}$, $T^0_f$ = 8.66K and $z\nu$=11.1. 
Here, the value of relaxation time constant  $\tau_0 = 3.6 \times 10^{-6}$ is larger at least by a factor of $10^3$ than the normal value for spin glass phase. The reason for this is not understood. However, the value of $T^0_f (=8.7K)$ is close to the $T^0_f$ value 10.1K obtained from AT line analysis. The obtained value of critical exponent `$z\nu$' (11.1K) matches well with the typical value of `$z\nu$' for spin-glass system ranging from 4 to 12 \cite{mydosh,bedanta}, suggesting a spin-glass ground state in $NiCr_2O_4$ nanoparticles.

The second approach assumes that the freezing phase transition is a non-equilibrium phenomenon and the dynamical properties of spin glass phase can be explored by Vogel-Fulcher law. This law takes into account the interacting property of spin-glass clusters. The Vogel-Fulcher law is expressed as,

\begin{equation}
\tau = \tau_0 exp\left[\frac{E_a}{k_{B}(T_f-T_0)}\right],
\label{vf law}
\end{equation}

here, $E_a$ represents the energy barrier or activation energy and $T_0$  is the phenomenological parameter describing interaction between clusters. The plot of $T_f$ vs. ln(1/f) is plotted together with the fitted curve given in eq. \ref{vf law} and is shown in the inset (b) of Fig. \ref{chi}. The fitted parameters are $E_a/k_B$ = 58.9K, $\tau_0 = 5.22 \times 10^{-8}$ and $T_0$ = 8.03K. The non-negative value of $T_0$ = $8.03$K indicates the presence of interacting spin-glass clusters. Moreover, for $T_0=0$, eq. \ref{vf law} reduces to Arrhenius law given by $\tau=\tau_0 \hspace{0.1cm} exp(E_a/k_B T_f)$. In this situation, the inter-cluster interaction is negligible. In view of this relation, the fitted data give unreasonably low value of $\tau_0$ ($\sim 10^{-40}$sec), which seems much faster than the atomic relaxation of spins ($\sim 10^{-14}$sec) which is physically not possible.


Now our basic concern is to understand the mechanism responsible for the generation of spin glass like phase in $NiCr_{2}O_{4}$ nanoparticle which shows a different behaviour as compared to its bulk counterpart. The 2D nature of the spin glass phase established using AT-line fitting indicates a surface effect. This may be associated with either spin canting or surface spin disorder. Spin canting exists due to two competing magnetic exchanges (one is isotropic and the other is asymmetric in nature). The marginal weakening of ferrimagnetic transition in 75h sample is an indication of weakening of isotropic exchange. Further, the bulk $NiCr_{2}O_{4}$ does not show glassy feature in our measurements confirming existing reports to the best of our knowledge. Therefore, we ruled out spin canting as the reason for existing surface effect in $NiCr_{2}O_{4}$ nanoparticle. Actually, surface effects result from the lack of translational symmetry at the boundaries of the particle because of the lower coordination number there and the existence of broken magnetic exchange bond which leads to surface spin disorder and frustration \cite{batlle}. Another important effect is the finite size effect which is observed in nanoparticle sample. This effect originates from the cut off of some characteristic length due to the purely geometric constraint on finite volume. It results in superparamagnetic behaviour, which is not present in our sample (already pointed out in earlier discussion). Finally, it is observed that nanoparticle sample has higher value of mean field frustration parameter `f' (23.6) as compare to its bulk sample (f=20). Also, weakening of $T_C$ in nanoparticle sample indicates absence of long range ferrimagnetic ordering and in fact exhibits spin-glass behaviour. This fulfils the general condition for the system to be geometrically frustrated magnet (GFM). Hence, the $NiCr_{2}O_{4}$ nanoparticle sample might be retaining geometrical frustration.

\section{Conclusion} 
From the entire study, the following points can be concluded:
1. The nanoparticle of $NiCr_{2}O_{4}$ shows 2D spin glass like character at low temperature with $T_f$= 10.1K. Vogel-Fulcher law analysis confirms the existence of interacting spin clusters.\\
2. The surface spin disorder seems to be the reason for the existence of 2D spin glass feature.\\
3. The decrease in magnetization value of $NiCr_{2}O_{4}$ nanoparticle compared to its bulk counterpart may be due to the surface spin disorder.\\
4. The existence of new transition ($T_P$) may be due to the presence of weak ferrimagnetic spin clusters.\\
5. The $NiCr_{2}O_{4}$ nanoparticles may be considered as a  GFM due to higher value of `f' parameter, absence of long range ordering and the presence of spin glass ground state.

\section{Acknowledgement} We would like to thank Ministry of Human Resource and Development (MHRD), Govt. of India, for funding. 



\begin{thebibliography}{}
%
%
\bibitem{krupika} S. Krupika, and P. Novak, (\textit{Ferromagnetic Materials}), edited by E. P. Wolfarth, Vol 3, p.189. North-Holand, Amsterdam (1982).
\bibitem{tomiyasu_77} K. Tomiyasu, H. Hiraka, K. Ohoyama, and K. Yamada, Resonance-Like Magnetic Excitations in Spinel Ferrimagnets $FeCr_2O_4$ and $NiCr_2O_4$ Observed by Neutron Scattering, J. Phys. Soc. Jpn., 77, 12 (2008).
\bibitem{klemme} S. Klemme, and J. C. Miltenburg, Thermodynamic properties of nickel chromite (NiCr2O4) based on adiabatic calorimetry at low temperatures, Phys. Chem. Miner., 29, 663 (2002).
\bibitem{crottaz} O. Crottaz, F. Kubel, and H. Schmid, Jumping crystals of the spinels $NiCr_2O_4$ and $CuCr_2O_4$, J. Mater. Chem., 7, 143 (1997).
\bibitem{mufti} N. Mufti, A. A. Nugroho, G. R. Blake, and T. T. M. Palstra, Magnetodielectric coupling in frustrated
spin systems: the spinels $MCr_2O_4$ (M = Mn, Co and Ni), J. Phys. Cond. Mat., 22, 075902-07 (2002).
\bibitem{tomiyasu} K. Tomiyasu, and I. Kagomiya, Magnetic Structure of $NiCr_2O_4$ Studied by Neutron Scattering and Magnetization Measurements, J. Phys. Soc. Jpn., 73, 2539-2542 (2004).
\bibitem{tejada1} J. Tejada, R. F. Ziolo, and X. X. Zhang, Quantum Tunneling of Magnetization in Nanostructured Materials, Chem. Mater., 8, 1784 (1996).
\bibitem{tejada2} J. Tejada, X. X. Zhang, E. del Barco, J. M. Hernandez, and E. M. Chudnovsky, Macroscopic Resonant Tunneling of Magnetization in Ferritin, Phys. Rev. Lett., 79, 1754 (1997).
\bibitem{kneller} E. F. Kneller, and F. E. Luborsky, Particle Size Dependence of Coercivity and Remanence of Single Domain Particles, J. Appl. Phys., 34, 656 (1963).
\bibitem{prince} E. Prince, Structure of Nickle Chromite*, J. Appl. Phys., 32, 68S (1961).
\bibitem{ishibashi} H. Ishibashi, and T. Yasumi, Structural transition of spinel compound $NiCr_2O_4$ at ferrimagnetic
transition temperature, J. Magn. Magn. Mater., 310, e610-e612 (2007).
\bibitem{coey} J. M. D. Coey, Noncollinear Spin Arrangement in Ultrafine Ferrimagnetic Crystallites, Phys. Rev. Lett., 27, 1140 (1971).
\bibitem{morrish} A. H. Morrish, K. Hanada, and P. J. Schurer, J. Phys. (Paris), Colloq. \textbf{37}, C6 (1976).
\bibitem{parker} F. T. Parker, M. W. Foster, D. T. Margulies, and  A. E. Berkowitz, Spin canting, surface magnetization, and finite-size effects in $\gamma-Fe_2O_3$ particles, Phys. Rev. B, 47, 7885 (1993).
\bibitem{ueda} H. Ueda, H. Mitamura, T. Goto, and Y. Ueda, Successive field-induced transitions in a frustrated antiferromagnet $HgCr_2O_4$, Phys. Rev. B., 73, 094415 (2006).
\bibitem{chen} X. H. Chen, H. T. Zhang, C. H. Wang, X. G. Luo, and P. H. Li, Effect of particle size on magnetic properties of zinc chromite synthesized by sol–gel method, Appl. Phys. Lett., 81, 4419 (2009).
\bibitem{kodama} R. H. Kodama, A. E. Berkowitz, E. J. McNiff, Jr., and S. Foner, Surface Spin Disorder in $NiFe_2O_4$ Nanoparticles, Phys. Rev. Lett., 77, 394 (1996). 
\bibitem{ramirez} K. H. J. Buschow, \textit{Handbook of magnetic material}, edited by K. H. J. Buschow, p-423., Vol. 13, North Holland, Amsterdam (2001).
\bibitem{at} J. R. L. Almeida, and D. J. Thouless, Stability of the Sherrington-Kirkpatrick solution of a spin glass model, J. Phys. A, 11, 983 (1978).
\bibitem{mydosh} J. A. Mydosh, Spin glasses: An Experimental Introduction, Taylor and Francis, London (1993).
\bibitem{souletie} J. Souletie and J. L. Tholence, Critical slowing down in spin glasses and other glasses: Fulcher versus power law, Phys. Rev. B, 32, 516(R) (1985).
\bibitem{edwards} E. A. Edwards and P. W. Anderson, Theory of spin glasses, J. Phys. F: Met. Phys., 5, 965 (1975).
\bibitem{mukadam}M. D. Mukadam, S. M. Yusuf, P. Sharma, S. K. Kulshreshtha, and G. K. Dey, Dynamics of spin clusters in amorphous $Fe_2O_3$, Phys. Rev. B, 72, 174408 (2005).
\bibitem{bedanta} S. Bedanta, and W. Kleemann, Supermagnetism, J. Phys. D: Appl. Phys., 42, 013001 (2009). 
\bibitem{nam} D. N. H. Nam, R. Mathieu, P. Nordblad, N. V. Khiem, and N. X. Phuc, Spin-glass dynamics of $La_{0.95}Sr_{0.05}CoO_3$, Phys. Rev. B, 62, 8989 (2000).
\bibitem{gunnarsson} K. Gunnarsson, P. Svedlindh, P. Nordblad, L. Lundgren, H. Aruga, and A. Ito, Dynamics of an Ising Spin-Glass in the Vicinity of the Spin-Glass Temperature, Phys. Rev. Lett., 61, 754 (1988).
\bibitem{batlle} X. Batlle, and A. Labarta, Finite-size effects in fine particles: magnetic and transport properties, J. Phys. D: Appl. Phys., 35, R15-R42 (2002).
\end{thebibliography}


\end{document}